\documentclass[12pt,a4]{revtex4}
\usepackage[T1]{fontenc}
\usepackage{times}
\usepackage{latexsym} 
\usepackage{amsfonts, amsthm, amsmath,amssymb}
\usepackage{times}
\usepackage{color}
\usepackage{xcolor}
\usepackage{mdframed}
\definecolor{refkey}{rgb}{0,0,1}
\definecolor{labelkey}{rgb}{1,0,0}

\def\<{{\langle}} 
\def\>{{\rangle}}

\def\note#1{{}}

\def\note#1{} 

\def\beq{\begin{equation}} 
\def\eeq{\end{equation}}

\def\id{\mathrm{id}}

\newcounter{zlist} 

\newcounter{blist} 
 
\newcounter{rlist} 
 
{ 
 \def\stac#1{\raise-.2cm\hbox{$\stackrel{\displaystyle\otimes}{\scriptscriptstyle{#1}}$}}

\def\cten#1{\raise-.2cm\hbox{$\stackrel{\displaystyle\widehat{\otimes}}{\scriptscriptstyle{#1}}$}}

\def\Label#1{\label{#1}\ifmmode\llap{[#1] }\else 
\marginpar{\smash{\hbox{\tiny [#1]}}}\fi} 
\def\Label{\label} 

\theoremstyle{definition}

\theoremstyle{remark}

\newcounter{c} 
 
\newcommand{\etyk}[1]{\vspace{-7.4mm}$$\begin{equation}\Label{#1} 
\addtocounter{c}{1}} 
\renewcommand{\]}{\ifnum \value{c}=1 $$\else \end{equation}\fi} 
\setcounter{tocdepth}{2} 

\def\CC{{\mathbb C}}

\def\HH{{\mathbb H}}

\def\ZZ{{\mathbb Z}}

\newcommand{\Aa}{\mathcal{A}}

\newcommand{\Cc}{\mathcal{C}}

\newcommand{\Jj}{\mathcal{J}}

\def\*C{{}^*\hspace*{-1pt}{\Cc}}
\def\text#1{{\rm {\rm #1}}}

\def\1{\mathbf{1}}

\newcounter{mnotecount}[section]
\renewcommand{\themnotecount}{\thesection.\arabic{mnotecount}}
\newcommand{\mnote}[1]
{\protect{\stepcounter{mnotecount}}$^{\mbox{\footnotesize
$
\bullet$\themnotecount}}$ \marginpar{
\raggedright \tiny\em
$\bullet$\themnotecount: #1} }

\textheight 22.8 cm
\textwidth 15.6cm
\topmargin -.25in \headheight 0.3in \headsep .5cm
\oddsidemargin .15in \evensidemargin .15in
\topskip 12pt
\begin{document} 
\vspace*{-2cm}
\title{A spectral geometry for the Standard Model without the fermion doubling.} 
\author{Arkadiusz Bochniak}
\author{Andrzej Sitarz}
\affiliation{Institute of Physics, Jagiellonian University,
	\hbox{prof.\ Stanis\l awa \L ojasiewicza 11, 30-348 Krak\'ow, Poland.}}

\begin{abstract} 
We propose a simple model of noncommutative geometry to describe the structure
of the Standard Model, which satisfies spin${}_c$ condition, has no 
fermion doubling, does not lead to the possibility of color symmetry 
breaking and explains the CP-violation as the failure of the reality condition 
for the Dirac operator.
\end{abstract} 
\maketitle 
\vspace*{-1cm}

\section{Introduction}

The Standard Model of particle interactions is certainly one of the most successful
and best tested theory about the fundamental constituents of matter and the
forces between them. Even though we still have no satisfactory description of
the strong interactions in low-energy regime and there are some puzzles concerning
masses and character of neutrinos as well as there are some experimental signs
that could point out to new physics, the Standard Model appears to be robust
and verified. Yet neither the content of fermion sector, the mixing between
the families and the fundamentally different character of the Higgs boson from
other gauge bosons appear to have a satisfactory geometrical explanation.

One of the few theories that aimed to provide a sound geometrical basis for
the structure of the Standard Model, explaining the appearance of the Higgs
and symmetry-breaking potential, was noncommutative geometry (see \cite{Co95,Co96,CCM07}). 
Constructed with the core idea that spaces with 
points can be replaced with algebras provided a plausible explanation of the gauge 
group of the 
Standard Model and the particles in its representation as linked to the unitary
group of a finite-dimensional algebra. Merged with the Kaluza-Klein idea that the 
physical spacetime has extra dimensions, the geometry of the finite-dimensional
algebra  (in the noncommutative sense) gave rise to the Higgs field understood 
as a connection, and the Higgs symmetry-breaking potential appeared as the 
usual Yang-Mills term in the action. 

The original model, which is based on the construction of a product geometry,
with the resulting geometry being the tensor product of a usual "commutative"
space with the finite-dimensional noncommutative geometry suffers from 
two problems. Firstly, in the original formulation it is Euclidean. Secondly, the
product structure leads to the quadrupling of the degrees of freedom in the
classical Lagrangian \cite{GBIS98,DKL16}. Moreover, the conditions put on the Dirac operator
for the finite geometry are not sufficient to restrict the class of possible operators to the physical one, leaving the possibility for the non-physical $SU(3)$-breaking 
geometries \cite{PSS99,DDS18, BS18, FB14a}. Though the latter problems appear to have at 
least a partial solution \cite{BS18} we believe that they can  be completely avoided if the 
noncommutative geometry behind the Standard Model is assumed to be 
spin${}_c$ only. 

In what follows we present a spin${}_c$ description of the geometry for
the Standard Model, which does not require fermion doubling, satisfies
the spin${}_c$ duality for spinors provided that the mass matrices and mixing
matrices are non-degenerate. The crucial role is then played then not by the 
Lorentzian Dirac operator but rather by its Krein-shift $\widetilde{D}$, 
the product of the Krein space fundamental symmetry $\beta$ and the Dirac 
operator $D$. This operator can be understood as the selfadjoint component 
of the Krein decomposition of the Lorentzian Dirac operator, $D = \beta \widetilde{D}$. 
Moreover, we link the breaking of the $J$-condition between the real structure and the Dirac 
operator to the appearance of the CP-symmetry breaking in the Standard Model.

\section{Dirac operator for the Standard Model}
The Dirac operator for the four-dimensional Minkowski space is of the form
$D = i \gamma^\mu \partial_\mu$,
with the gamma matrices satisfying the relation 
$\gamma^\mu \gamma^\nu + \gamma^\nu \gamma^\mu = 2 \eta^{\mu\nu}$, 
where $\eta^{\mu\nu}$ is the standard Minkowski metric of signature 
$(+,-,-,-)$. 
We use the conventions of \cite{BS18}, so that $\gamma^0$
is selfadjoint and the remaining gamma matrices are 
antiselfadjojnt.

The Lorentz-invariant fermionic action, which leads to the Dirac equation, is
\begin{equation}
\int_M \overline{\psi} D \psi = \int_M \psi^\dagger \widetilde{D} \psi,
\label{fact}
\end{equation}
where $\overline{\psi} = \psi^\dagger \gamma^0$ and $\widetilde{D} = \gamma^0 D$.
The operator, $\widetilde{D}$ is a symmetric operator, which we call Krein-shift of the Dirac operator. This follows
from the properties of the Lorentzian Dirac operator $D$, which is  Krein-selfadjoint 
\cite{PaSi}, $D^\dagger = \gamma^0 D \gamma^0$, where $\gamma^0$ is the fundamental symmetry
of the Krein space. Written explicitly in the chiral  representation it becomes
\begin{equation}
\widetilde{D} = 
i\begin{pmatrix}
\sigma^\mu & 0 \\
0 & \widetilde{\sigma}_\mu
\end{pmatrix}\partial_\mu,
\end{equation}
where $\sigma^\mu$ and $\tilde{\sigma}^\mu$ are the standard and associated Pauli matrices, 
$\widetilde{\sigma}^0=\sigma^0, \widetilde{\sigma}^k=-\sigma^k$.

The Lorentzian Dirac operator and the related Lorentzian spectral triple have the standard $\ZZ_2$-grading $\gamma$ 
and the charge conjugation operator given,
\begin{equation}
\gamma = 
\begin{pmatrix}
1_2 & 0 \\
0 & -1_2
\end{pmatrix}, 
\qquad
\mathcal{J} =i\gamma^2\circ cc
= i \begin{pmatrix}
0 & \sigma^2\\
-\sigma^2 & 0
\end{pmatrix} \circ cc,
\end{equation}
where $cc$ denotes the usual complex conjugation of spinors. The operators $D,\gamma,\mathcal{J}$ satisfy the usual commutation relations for the 
geometry of the signature $(1,3)$:
\begin{equation}
D\gamma=-\gamma D, \qquad D\mathcal{J}=\mathcal{J}D,\qquad \mathcal{J}^2=1,\qquad \mathcal{J}\gamma=-\gamma\mathcal{J},
\end{equation}
whereas for the Krein-shifted operator we have
\begin{equation}
{\widetilde{D}\gamma=\gamma\widetilde{D}}, \qquad \widetilde{D}\mathcal{J}=-\mathcal{J}\widetilde{D}, \qquad \mathcal{J}^2=1, \qquad \mathcal{J}\gamma=-\gamma\mathcal{J}.
\end{equation}

The so-far accepted and tested experimentally action for the Standard Model of fundamental
interactions can be viewed as the extension of the action for a single bispinor to a family
of particles, with the additional terms in the action arising from a slight modification of
the Dirac operator by an endomorphism of the finite-dimensional space of fermions.

Before we discuss this extension and the conditions it satisfies we recall the notion of
Riemannian spectral triples and spin${}_c$-spectral triples, which are a bigger class than 
these arising from generalisation of the spin geometry only.

\section{Riemannian and pseudoriemannian spectral triples}
A Riemannian finite spectral triple \cite{LRV12} built over a finite-dimensional 
algebra $A$ is a collection of data $(A, D, H, \pi_L, \pi_R)$, where $\pi_L$ is the representation
of $A$ on $H$, $\pi_R$ is the representation of $A^{op}$ (the opposite algebra 
to $A$) on $H$ such that:
\begin{equation}
\left[ \pi_L(a), \pi_R(b) \right] = 0,
\end{equation}
\begin{equation}
\label{first_order}
\left[ [D, \pi_L(a)], \pi_R(b) \right] = 0,
\end{equation}
for all $a\in A$ and $b\in A^{op}$.

We say that the spectral triple is of spin${}_c$ (see \cite{DD16} and compare
with the classical result \cite{Pl86}) type if 
\begin{equation} (Cl_D(\pi_L(A))' = \pi_R(A),\end{equation}
or of Hodge type if 
\begin{equation}(Cl_D(\pi_L(A))' = Cl_D(\pi_R(A)). \end{equation}
By the generalized Clifford algebra $Cl_D(\pi_L(A))$ (and similarly $Cl_D(\pi_R(A))$) we understand the 
algebra generated by $\pi_L(a)$ and $[D,\pi_L(b)]$ for all $a,b \!\in\!A$.

Of course, genuine Riemannian geometries require further assumption that  the
operator $D$ has a compact resolvent. In the case of Lorentzian or, more generally,
pseudoriemannian geometries, we might follow the path of \cite{PaSi} extending the
definition of Lorentzian real spectral triples to Lorentzian spin${}_c$ geometries.

\section{Fermions and the algebra of the Standard Model}

Let us recall a convenient parametrization of the particle content in the one-generation Standard Model \cite{DD16}:
\begin{equation}
\Psi=\begin{pmatrix}
\nu_R & u_R^1 & u_R^2 & u_R^3 \\
 e_R & d_R^1 & d_R^2 & d_R^3 \\
\nu_L & u_L^1 & u_L^2 & u_L^3 \\
 e_L & d_L^1 & d_L^2 & d_L^3
\end{pmatrix} \in M_4(H_W),
\end{equation}
where each of the entries is the Weyl spinor over the Minkowski space with a fixed 
chirality. As the algebra $\Aa$ we take the algebra of functions over the Minkowski space,  valued in $ \CC \oplus \HH \oplus M_3(\CC)$ and chose the following two representations of the algebra: 
$$ 
\pi_L(\lambda, q ,m ) \Psi = 
\left( \begin{array}{ccc} \lambda & & \\ & \bar{\lambda} & \\ & & q \end{array}\right) \Psi ,
\qquad
\pi_R(\lambda, q ,m ) \Psi = 
\Psi \begin{pmatrix}
\lambda &  \\
& m^T
\end{pmatrix}.
$$
where $\lambda, q, m$ are complex, quaternion and $M_3(\CC)$-valued functions, respectively. The representation $\pi_L$ acts by multiplying $\Psi$ from the left whereas $\pi_R$ acts by multiplying $\Psi$ from the right.  This is the reason that we transpose 
$m$ so that $\pi_R$ is indeed a representation.  Observe that since left and right multiplication commute then $[\pi_L(a),\pi_R(b)]=0$ for all $a,b\in \Aa$, i.e. the zero-order 
condition is satisfied.  Due to the simplicity of the notation at every point of the Minkowski space we can encode any linear operator on the space of particles as an operator in $M_4(\CC) \otimes M_2(\CC)\otimes M_4(\CC)$, where the the first and the last matrix act by multiplication 
from the left and from the right and the middle $M_2(\CC)$ matrix acts on the 
components of the Weyl spinor.

The full Lorentzian Dirac operator of the Standard Model is, in this notation, of the form
\begin{equation}
D_{SM} \Psi = \underbrace{
\begin{pmatrix}
&& i\widetilde{\sigma}^\mu\partial_\mu &\\
&&& i\widetilde{\sigma}^\mu \partial_\mu \\
i\sigma^\mu \partial_\mu &&&\\
& i \sigma^\mu \partial_\mu &&
\end{pmatrix}}_{D} \Psi + D_F\Psi,
\end{equation}
where $D_F$ is a finite endomorphism of the Hilbert space $M_4(H_W)$.

First of all, observe that the spatial part $D$ is covariant under the Lorentz 
transformations so that the Lagrange density (\ref{fact}) is invariant. 
Indeed, using the $SL(2,\CC)$ representation of the Lorentz group with an appropriate transformation of the Weyl spinors, 
it is obvious that $D$ transforms covariantly. On the other hand, $D_F$ will transform covariantly, so that the full fermionic action will remain invariant under Lorentz transformations, only if it is an element of $M_4(\CC) \otimes \id \otimes M_4(\CC)$, 
so it is a scalar  from the point of view of Lorentz transformations. 

At this point it is the Lorentz invariance and the requirement that $D_F$ behaves
like a scalar under Lorentz transformations that fixes $D_F$ to commute with the 
chirality $\Gamma$, which, in fact, can be written as an element of the algebra of the Standard Model, $\Gamma = \pi_L(1,-1,1)$. In the end we have the genuine Lorentzian
Dirac operator $D$ that anticommutes with $\Gamma$ and the finite part of the
full Dirac operator, $D_F$, commuting with $\Gamma$, whereas the Krein-shifted
parts have the opposite behavior. 

Next, we see what are the sufficient conditions that the Krein-shifted operator $\widetilde{D_{SM}}$ satisfies the first order condition for the given 
algebra and the representation. First, observe that $\widetilde{D}$ alone obviously 
satisfies the order one condition and therefore we need to check only 
$\widetilde{D_F}$. Suppose then that
$$ [[\widetilde{D_F},\pi_L(a)],\pi_R(b)]=0,$$
for all $a,b\in \mathcal{A}$. As any element in $\pi_L(\Aa)$ commutes with $\pi_R(\Aa)$
it suffices to find all $\widetilde{D_F}$ that are selfadjoint, commute with the elements from $\pi_R(\Aa)$ and anticommute with 
$\Gamma$. It is easy to see that such operators are restricted to
\begin{equation}
\widetilde{D_F} = 
\underbrace{\begin{pmatrix}
& M_l \\ M_l^\dagger &
\end{pmatrix}}_{D_l} \otimes e_{11} +
\underbrace{\begin{pmatrix}
	& M_q \\ M_q^\dagger &
\end{pmatrix}}_{D_q} \otimes (1_4 - e_{11}),
\end{equation}
where $M_l,M_q\in M_2(\mathbb{C})$.

\subsection{The spin${}_c$ condition}

The Krein-shifted Dirac operator satisfies first order condition, yet it still may not 
provide the spin${}_c$ spectral geometry.  We shall look for necessary and sufficient 
conditions that the commutant of the (complexified) Clifford algebra,  $Cl_D(\pi_L(\Aa))$, generated 
by $\pi_L(\Aa)$ and $[\widetilde{D_{SM}}, \pi_L(\Aa)]$ is $\pi_R(\Aa)$. First, observe that 
all operators in the so defined $Cl_D(\pi_L(\Aa))$ are endomorphisms of the space $M_4(H_W)$, which contain a subalgebra generated by the commutators of $\widetilde{D}$  with functions $C^\infty_{\mathbb{C}}(M)$. This subalgebra, acts on the Weyl spinors pointwise and can be identified with $M_2(\CC) \oplus M_2(\CC)$-valued functions on the Minkowski space. The resulting subalgebra of the Clifford algebra acts only on the Weyl-spinorial components, separately in the left and in the right sector. The commutant of this algebra in the endomorphisms of the Hilbert $M_4(H_W)$ space is then contained in the $M_4(\CC) \otimes \id \otimes  M_4(\CC)$ (at each point of the Minkowski space).

Further, consider the subalgebra generated by the commutators of $\widetilde{D_{F}}$ 
with constant functions in $\Aa$. It is a subalgebra of $M_4(\CC) \otimes \id
\otimes (\CC \oplus \CC^{(3)})$-valued constant functions over the Minkowski space 
and it is easy to see that both subalgebras generate the full Clifford algebra. Therefore the common commutant of both parts will be the commutant of the full Clifford algebra. 

From the decomposition it is easy to see that the commutant of second part are the
functions in $\id \otimes M_2(\CC) \otimes (\CC \oplus M_3(\CC))$ and therefore, the common part are functions valued in $\id \otimes\id \otimes(\CC \oplus M_3(\CC))$, which
indeed is the algebra $\pi_R(\Aa)$.

\subsection{Three generations}

Let us consider three families of leptons and quarks, that is the Hilbert space
$M_4(H_W) \otimes \CC^3$ with the diagonal representation of the algebra. The
only difference from the previous section is that the matrices $M_l$ and $M_q$ are no longer in 
$M_2(\CC)$ but in $M_2(\CC) \otimes M_3(\CC)$. As the algebra acts diagonally on the Hilbert space 
(with respect to the generations) we can again repeat the arguments of \cite{DS18} and argue that the 
spin${}_c$ condition will hold if algebras generated by $\pi_L(\Aa)$ and $D_l,D_q$, respectively, 
will be full matrix algebras, that is $(M_4(\CC) \otimes \id \otimes \id) \otimes M_3(\CC)$, independently
for lepton and for quarks. 

Since the arguments we have used here are analogous to ones used in the discussion of {\it full conditions} (4.2.2 in \cite{DS18}), we infer the same condition for the Hodge property to be satisfied.

Both $M_l$ and $M_q$ can be diagonalized, yet because of the doublet structure
of the left leptons and quarks the components (up/down) cannot be diagonalized
simultaneously. The standard presentation of the mass matrices for the Physical
Standard Model is then

$$ 
M_l = \begin{pmatrix} \Upsilon_\nu & 0 \\ 0 & \Upsilon_e \end{pmatrix},
\qquad
M_q= \begin{pmatrix} \Upsilon_u & 0 \\ 0 & \Upsilon_d \end{pmatrix},
$$
where $\Upsilon_e$ and $\Upsilon_u$ are chosen diagonal with the masses 
of electron, muon, tau and the up, charm, top quarks, respectively and
$$  
\Upsilon_\nu = U \widetilde{ \Upsilon_\nu} U^\dagger,  \qquad
\Upsilon_d = V \widetilde{ \Upsilon_d} V^\dagger, 
$$
with diagonal matrices $\widetilde{ \Upsilon_\nu}, \widetilde{ \Upsilon_d}$ providing
(Dirac) masses of all neutrinos and down, strange, bottom quarks, where $U$ is
the Pontecorvo-Maki-Nakagawa-Sakata mixing matrix (PMNS matrix) and
$V$ is the Cabibbo-Kobayashi-Maskawa mixing matrix (CKM matrix).

As was indicated also in \cite{DS18} the sufficient condition to fulfill the Hodge property is that for both pairs of matrices $(\Upsilon_\nu,\Upsilon_e)$ and $(\Upsilon_e,\Upsilon_d)$ their eigenvalues are pairwise different. This requirement is satisfied in the case of the Physical Standard Model, provided that there is no massless neutrino (see 5.2 in \cite{DS18}).  

\subsection{From spin${}_c$ to Hodge condition}

Consider for a while the Hilbert space $H_{SM}=M_4(\CC)$ with the same left and right
representation of the algebra as in the Standard Model case (the SM Hilbert space
is the tensor product of the above with the space of Weyl fermions). Taken with the Krein-shifted Dirac $\widetilde{D_F}$ operator and $\Gamma = \pi_L(1,-1,1)$ it is a 
Euclidean even spectral triple.

Assume now that $\widetilde{D_F}$ is such that the spin${}_c$ condition holds. We
shall describe now the procedure of the {\em doubling} of the triple, so that the
resulting real spectral triple satisfies the Hodge duality and is the 
{\em finite spectral triple of the Standard Model} studied so far as the finite 
component of the product geometry.

Consider $H_{SM}^2 = H_{SM} \oplus H_{SM}$ with the representation $\pi_L \oplus \pi_R$.
We define the real structure $J$ as the composition of the hermitian conjugation with
the $\ZZ_2$ action exchanging the two copies of $H_{SM}$, so that 
$J (M_1 \oplus M_2) = M_2^* \oplus M_1^*$. 

It is clear that the conjugation by $J$ maps the representation of the algebra $A$
to its commutant. We extend $\Gamma$ so that the relation $J\Gamma=\Gamma J$
holds and extend the Dirac operator $\widetilde{D_F}$ in the following way, 
$$ D' = \widetilde{D_{F}} \oplus 0  + J (\widetilde{D_{F}} \oplus 0 ) J^{-1}. $$
Clearly $D'$ anticommutes with $\Gamma$ and commutes with $J$. The Clifford
algebra, that is the algebra generated by $\pi_L \oplus \pi_R$ and the commutators
with $D'$ is $Cl_{\widetilde{D_{F}} }(\pi_L(A)) \oplus \pi_R(A)$. Due to the fact that before the doubling we had the spin${}_c$-condition it is clear that the commutant of the Clifford algebra contains
$\pi_R (A)\oplus Cl_{\widetilde{D_{F}}}(\pi_L(A))$. It is therefore sufficient to verify that there are no other operators $T$ that map $H_{SM}$ to $H_{SM}$, which would satisfy that they 
commute with the representation of $Cl_{\widetilde{D_{F}} } (\pi_L(A))\oplus \pi_R (A)$.
Identifying the Hilbert space as $\CC^{16} \oplus \CC^{16}$ we see
that the first component of Clifford algebra is $M_4(\CC) \oplus M_4(\CC)^{(3)}$ (action
diagonally on $\CC^{16}$, the notation $B^{(n)}$ means that we take $n$-copies of the
algebra $B$) and the second is $\CC^{(4)} \oplus M_3(\CC)^{(4)}$. Since all these algebras are independent of each other there exists no operator intertwining their actions,
hence the commutant is exactly the one indicated above.

\subsection{The reality and the CP-violation}

Let us take the real structure $J$ acting on finite part just by the complex conjugation,
that is, the real structure implemented on $M_4(H_W)$ simply as 
$\id \otimes \Jj \otimes \id$. Of course, it does not implement the usual zero-order 
condition, however, we still have a milder version of the zero-order condition in the 
following form:
$$ \pi_R(\Aa) \subset J \pi_L(\Aa) J^{-1} . $$

We have already observed what are the commutation relations between $\widetilde{D}$
and $\Jj$ (and hence $J$). Next let us see whether similar commutation relations
can be imposed on $\widetilde{D_{F}}$. As both $J^2$ as well as the anticommutation
with $\Gamma$ are fixed, we see that by imposing the same KO-dimension (6) for the
Euclidean finite spectral triple as for the Lorentzian spatial part we shall have $J \widetilde{D_{F}} = \widetilde{D_{F}} J$. This condition is very mild and means that the 
mass matrices $M_l$ and $M_q$ have to be real. In case of one generation of particles 
it implies that masses of fermions have to be real, which is hardly very restrictive.
 
Yet the situation changes when we pass to three generations as already discussed
above when considering the spin${}_c$-condition. Since $J$ acts by complex conjugation
then the requirement $\widetilde{D_F} J = J \widetilde{D_F}$ still is equivalent to the
matrices $M_l,M_q$ having only real entries. Using the standard parametrization 
described above, this leads to the reality of the physical masses. However, since in the
case of three generations the matrices $\Upsilon_\nu, \Upsilon_d$ are not diagonal
we must ensure that both $U$ and $V$ mixing matrices as real. 
 
If this is the case, then all phases in the standard parametrization of these matrices should vanish, which physically will have the interpretation of the CP symmetry preservation.
However, in case of the CKM matrix it implies that the Wolfenstein parameter $\bar{\eta}$ has to vanish, but experimentally it is known that $\bar{\eta}=0.355^{+0.012}_{-0.011}$ \cite{PDG}.
The CP-violating phase $\delta_{\mathrm{CP}}^\nu$ in the neutrino sector, originated from the PMNS matrix, was determined to be $\delta_{\mathrm{CP}}^\nu /\pi={1.38}_{-0.38}^{+0.52}$ \cite{PDG, PDGnu}, that strongly confirms the CP symmetry breaking. Therefore, the existence of CP-violation may be interpreted as 
a shadow of $J$-symmetry violation in the non-doubled spectral triple. 

\subsection{Twisted (pseudoriemannian) spectral triple}

We have verified that the Krein-shifted Dirac operator satisfies the 
order-one condition (\ref{first_order}). It appears that this is equivalent to the 
Lorentzian Dirac operator $D_{ST} = \beta \widetilde{D_{ST}}$  satisfying a twisted 
version of the order-one condition, that is,
\begin{equation}
\left[ [D_{ST}, \pi_L(a)]_{\beta},\pi_R(b) \right]_{\beta}=0,
\end{equation}
where $[x,y]_{\beta}= x y - \beta y \beta^{-1} x$ and $\beta = \mathrm{id}\otimes\gamma^0 \otimes \id$. 
This follows directly from a simple computation, which uses $\beta^2=\id$:
$$
\begin{aligned}
0 =& \left[[{\widetilde{D_{ST}}},\pi_L(a)],\pi_R(b)\right] 
 =\beta D_{ST}\pi_L(a)\pi_R(b) -\pi_L(a)\beta D_{ST}\pi_R(b) \\ 
& -\pi_R(b)\beta D_{ST}\pi_L(a) +\pi_R(b)\pi_L(a)\beta D_{ST} 
 = \beta \left(  D_{ST}\pi_L(a)\pi_R(b) - \beta\pi_L(a)\beta D_{ST}\pi_R(b) \right. \\
&  \left. -\beta\pi_R(b)\beta D_{ST}\pi_L(a)+\beta\pi_R(b)\pi_L(a)\beta D_{ST} \right)
= \beta \left[ [D,\pi_L(a)]_\beta,\pi_R(b)\right]_\beta.
\end{aligned}
$$

\section{Conclusions}

Let us stress that the geometry of the Standard Model, as discussed above, is not 
a product of spectral triples. Nevertheless, it has interesting features, which
we summarize here with an outlook for the future research directions.

When restricted to the commutative algebra of real-valued functions (and its
complexification) we obtain the even Lorentzian spectral triple with a real structure,
so that the Dirac operator satisfies order-one condition and which is of 
KO-dimension $6$ (compatible with the signature $(1,3)$). 

On the other hand, the restriction of the spectral triple to the constant functions
over the Minkowski space gives a Euclidean even spectral triple, which, 
fails to be real. The failure of the real structure to satisfy the commutation relation 
with the (Krein-shifted) finite part of the Dirac operator is tantamount to the 
appearance of the violation of CP symmetry in the Standard Model.

Neither of the restrictions does satisfy the spin${}_c$-condition, as in both cases we
still consider the full Hilbert space. Yet the full spectral triple satisfies the spin${}_c$-condition in the following sense: the Clifford algebra generated by the 
commutators of the Krein-shifted Dirac operator with the representation $\pi_L$ of the algebra has, as the commutant, the right representation of the algebra
$\pi_R$.

There are several possible ramifications of the above observations. First is the
disappearance of the product structure: yet even if the triple is not a full product, 
then possibly it can have some structure of a quotient "spectral geometry". It
will be interesting to classify all possible covers and all Dirac operators
for them. In the presented spectral triple the family of allowed Dirac operators
that satisfy the spin${}_c$ condition is much closer to physical reality as it does
not include any color-symmetry breaking operator unlike \cite{DDS18} and, moreover,
the conditions are exactly the same as for the Hodge duality. The failure of the finite
spectral triple to be real is then a geometric interpretation of the CP-symmetry 
breaking in the Standard Model. Finally, the disappearance of the product structure
may have deep consequences for the spectral action. We postpone the discussion of 
possible effects on the physical parameters of the model for the forthcoming work.

\noindent {\bf Acknowledgements:} The authors thank L.D\k{a}browski for helpful comments.


\begin{thebibliography}{12}
	
\bibitem{Co95} A.~Connes, {\it Noncommutative Geometry and reality}, J. Math. Phys. {\bf 36} (1995), 6194-6231.
	
\bibitem{Co96} Connes A., {\it Gravity coupled with matter and foundation of non-commutative geometry}, 
Commun. Math. Phys. {\bf 182} (1996), 155-176.
	
\bibitem{CCM07}  A. H.~Chamseddine, A.~Connes, M.~Marcolli, {\it Gravity and the standard model with neutrino mixing} 
Adv. Theor. Math. Phys. {\bf 11}, 991-1089 (2007).

\bibitem{GBIS98}  
J.~M.~Gracia-Bondia, B.~Iochum and T.~Schucker,
{\it The Standard model in noncommutative geometry and fermion doubling},
Phys.\ Lett.\ B {\bf 416} (1998) 123

\bibitem{DKL16} F.~D'Andrea, M.~Kurkov, F.~Lizzi {\it Wick Rotation and Fermion Doubling in Noncommutative Geometry}, 
Phys. Rev. {\bf D 94}, 025030 (2016).


\bibitem{PSS99}
M.~Paschke, F.~Scheck and A.~Sitarz,
{\it Can (noncommutative) geometry accommodate leptoquarks?},
Phys.\ Rev.\ D {\bf 59} (1999) 035003


\bibitem{DDS18} L.~D\k{a}browski, F.~D'Andrea, A.~Sitarz, {\it The Standard Model in noncommutative geometry: fundamental fermions as internal forms}, Lett. Math. Phys. (2018) {\bf 108}: 1323.

\bibitem{BS18} A.~Bochniak, A.~Sitarz, {\it Finite pseudo-Riemannian spectral triples and the standard model}, Phys. Rev. {\bf D 97} (2018), 115029.	


\bibitem{FB14a}
S. Farnsworth and L. Boyle, \textit{Non-Commutative Geometry, Non-Associative Geometry and the Standard Model of Particle Physics}, New J. Phys. 16 (2014), 123027; 

\bibitem{PaSi}
M.~Paschke, A.~Sitarz, 
{\it Equivariant Lorentzian spectral triples}, arXiv preprint mathph/0611029

\bibitem{LRV12}
S. Lord, A. Rennie and J.C. V\'arilly, 
{\it Riemannian manifolds in noncommutative geometry}, 
J. Geom. Phys. 62 (2012) 1611-1638


\bibitem{DD16} L.~D\k{a}browski, F.~D'Andrea, {\it The Standard Model in Noncommutative Geometry and Morita equivalence}, J. Noncommut. Geom. 10, (2) 2016, 551-578.

\bibitem{Pl86}
R. Plymen, 
{\it Strong Morita equivalence, spinors and symplectic spinors}, 
J. Operator Theory 16(1986), no. 2, 305-324


\bibitem{DS18} L.~D\k{a}browski, A.~Sitarz, {\it Fermion masses, mass-mixing and the almost commutative geometry of the Standard Model}, J. High Energ. Phys. (2019) 2019: {\bf 68}.


\bibitem{PDG} Tanabashi M. et al. (Particle Data Group), Phys. Rev. {\bf D98}, 030001 (2018). 

\bibitem{PDGnu} Capozzi F. et al., {\it Global contraints on absolute neutrino masses and their ordering}, Phys. Rev. {\bf D95}, 096014 (2017).

\end{thebibliography}
\end{document}